\pdfoutput=1
\documentclass[twoside,twocolumn,9pt]{article}
\usepackage{extsizes}
\usepackage[super,sort&compress,comma]{natbib} 
\usepackage[version=3]{mhchem}
\usepackage[left=1.5cm, right=1.5cm, top=1.785cm, bottom=2.0cm]{geometry}
\usepackage{balance}
\usepackage{mathptmx}
\usepackage{sectsty}
\usepackage{graphicx} 
\usepackage{lastpage}
\usepackage[format=plain,justification=justified,singlelinecheck=false,font={stretch=1.125,small,sf},labelfont=bf,labelsep=space]{caption}
\usepackage{float}
\usepackage{fancyhdr}
\usepackage{fnpos}
\usepackage[english]{babel}
\addto{\captionsenglish}{%
  
}
\usepackage{array}
\usepackage{droidsans}
\usepackage{charter}
\usepackage[T1]{fontenc}
\usepackage[usenames,dvipsnames]{xcolor}
\usepackage{setspace}
\usepackage[compact]{titlesec}
\usepackage{hyperref}

\definecolor{cream}{RGB}{222,217,201}

\begin{document}

\pagestyle{fancy}
\thispagestyle{plain}
\fancypagestyle{plain}{
\renewcommand{\headrulewidth}{0pt}
}

\makeFNbottom
\makeatletter
\renewcommand\LARGE{\@setfontsize\LARGE{15pt}{17}}
\renewcommand\Large{\@setfontsize\Large{12pt}{14}}
\renewcommand\large{\@setfontsize\large{10pt}{12}}
\renewcommand\footnotesize{\@setfontsize\footnotesize{7pt}{10}}
\makeatother

\renewcommand{\thefootnote}{\fnsymbol{footnote}}
\renewcommand\footnoterule{\vspace*{1pt}%
\color{cream}\hrule width 3.5in height 0.4pt \color{black}\vspace*{5pt}} 
\setcounter{secnumdepth}{5}

\makeatletter 
\renewcommand\@biblabel[1]{#1}            
\renewcommand\@makefntext[1]%
{\noindent\makebox[0pt][r]{\@thefnmark\,}#1}
\makeatother 
\renewcommand{\figurename}{\small{Fig.}~}
\sectionfont{\sffamily\Large}
\subsectionfont{\normalsize}
\subsubsectionfont{\bf}
\setstretch{1.125} 
\setlength{\skip\footins}{0.8cm}
\setlength{\footnotesep}{0.25cm}
\setlength{\jot}{10pt}
\titlespacing*{\section}{0pt}{4pt}{4pt}
\titlespacing*{\subsection}{0pt}{15pt}{1pt}

\fancyfoot{}
\fancyfoot[LO,RE]{\vspace{-7.1pt}\includegraphics[height=9pt]{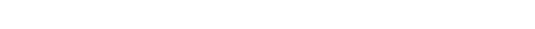}}
\fancyfoot[CO]{\vspace{-7.1pt}\hspace{13.2cm}\includegraphics{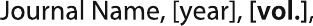}}
\fancyfoot[CE]{\vspace{-7.2pt}\hspace{-14.2cm}\includegraphics{head_foot/RF}}
\fancyfoot[RO]{\footnotesize{\sffamily{1--\pageref{LastPage} ~\textbar  \hspace{2pt}\thepage}}}
\fancyfoot[LE]{\footnotesize{\sffamily{\thepage~\textbar\hspace{3.45cm} 1--\pageref{LastPage}}}}
\fancyhead{}
\renewcommand{\headrulewidth}{0pt} 
\renewcommand{\footrulewidth}{0pt}
\setlength{\arrayrulewidth}{1pt}
\setlength{\columnsep}{6.5mm}
\setlength\bibsep{1pt}

\makeatletter 
\newlength{\figrulesep} 
\setlength{\figrulesep}{0.5\textfloatsep} 

\newcommand{\topfigrule}{\vspace*{-1pt}%
\noindent{\color{cream}\rule[-\figrulesep]{\columnwidth}{1.5pt}} }

\newcommand{\botfigrule}{\vspace*{-2pt}%
\noindent{\color{cream}\rule[\figrulesep]{\columnwidth}{1.5pt}} }

\newcommand{\dblfigrule}{\vspace*{-1pt}%
\noindent{\color{cream}\rule[-\figrulesep]{\textwidth}{1.5pt}} }

\makeatother

\twocolumn[
  \begin{@twocolumnfalse}
{\includegraphics[height=30pt]{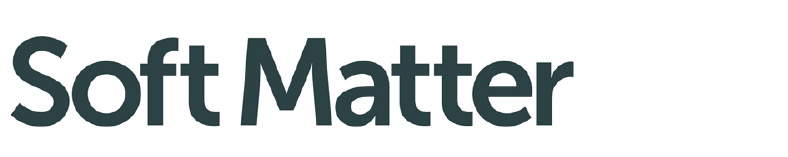}\hfill\raisebox{0pt}[0pt][0pt]{\includegraphics[height=55pt]{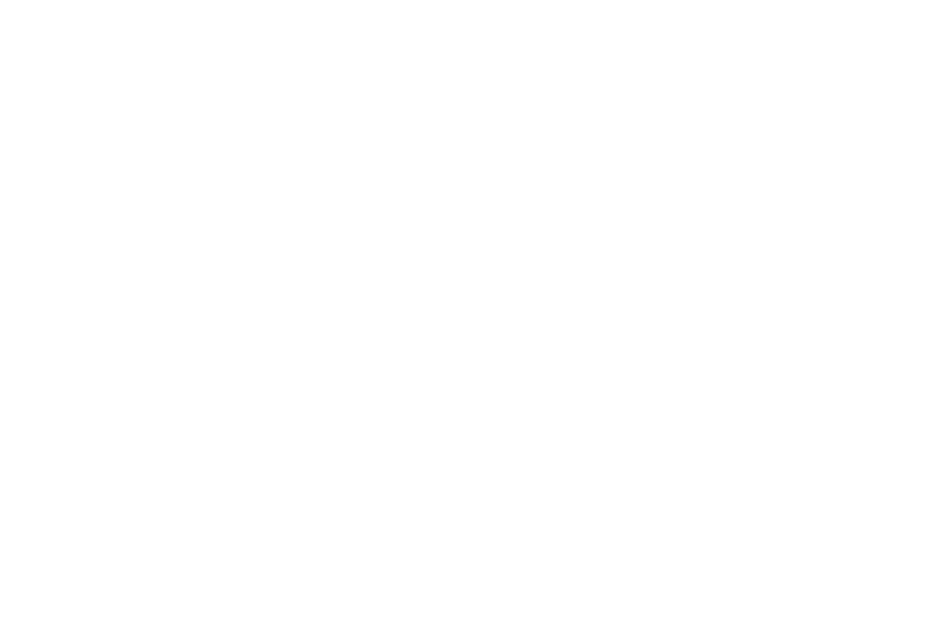}}\\[1ex]
\includegraphics[width=18.5cm]{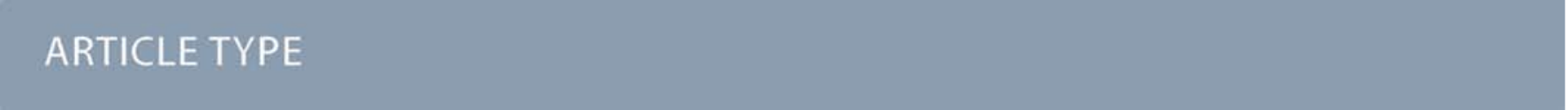}}\par
\vspace{1em}
\sffamily
\begin{tabular}{m{4.5cm} p{13.5cm} }

\includegraphics{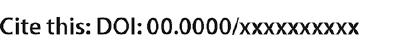} & \noindent\LARGE{\textbf{Flow and arrest in stressed granular materials$^\dag$}} \\
\vspace{0.3cm} & \vspace{0.3cm} \\

 & \noindent\large{Ishan Srivastava,$^{\ast}$\textit{$^{a}$} Leonardo E. Silbert,\textit{$^{b}$} Jeremy B. Lechman,\textit{$^{c}$} and Gary S. Grest\textit{$^{c}$}} \\

\includegraphics{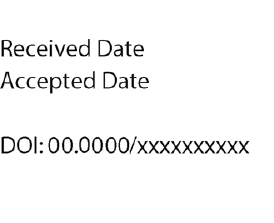} & \noindent\normalsize{Flowing granular materials often abruptly arrest if not driven by sufficient applied stresses. Such abrupt cessation of motion can be economically expensive in industrial materials handling and processing, and is significantly consequential in intermittent geophysical phenomena such as landslides and earthquakes. Using discrete element simulations, we calculate states of steady flow and arrest for granular materials under the conditions of constant applied pressure and shear stress, which are also most relevant in practice. Here the material can dilate or compact, and flow or arrest, in response to the applied stress. Our simulations highlight that under external stress, the intrinsic response of granular materials is characterized by uniquely-defined steady states of flow or arrest, which are highly sensitive to interparticle friction. While the flowing states can be equivalently characterized by volume fraction, coordination number or internal stress ratio, to characterize the states of shear arrest, one needs to also consider the structural anisotropy in the contact network. We highlight the role of dilation in the flow-arrest transition, and discuss our findings in the context of rheological transitions in granular materials.} \\

\end{tabular}

 \end{@twocolumnfalse} \vspace{0.6cm}

  ]

\renewcommand*\rmdefault{bch}\normalfont\upshape
\rmfamily
\section*{}
\vspace{-1cm}


\footnotetext{\textit{$^{a}$~Center for Computational Sciences and Engineering, Lawrence Berkeley National Laboratory, Berkeley, CA 94720, USA. E-mail: isriva@lbl.gov}}
\footnotetext{\textit{$^{b}$~School of Math, Science, and Engineering, Central New Mexico Community College, Albuquerque, NM 87106, USA. }}
\footnotetext{\textit{$^{c}$~Sandia National Laboratories, Albuquerque, NM 87185, USA. }}

\footnotetext{\dag~Electronic Supplementary Information (ESI) available: See DOI:}







\section{Introduction}
A remarkable property of granular materials is their ability to exist in both solid-like and fluid-like states.\cite{forterre2008} The fluid-like properties of these materials are commonly utilized during their production, handling and transportation in several industries such as pharmaceutical, agriculture and construction, while flowing granular materials are also observed in important geophysical phenomena such as flow of fault gouge in earthquakes and debris flow in landslides.\cite{jerolmack2019a} Frequently, however, the flowing granular material abruptly arrests leading to significant economic and geophysical consequences, such as clogged flows \cite{zuriguel2014} and cessation of sediment transport in riverbeds.\cite{clark2015} Such flow-arrest transitions are not limited to dry granular materials, but are also observed in dense suspensions, where the suspension viscosity dramatically increases by several orders of magnitude upon external stressing.\cite{brown2012,peters2016} Although the distinctions between solid-like and fluid-like states of granular materials have been thoroughly studied,\cite{jaeger1996} a unified understanding of the flow-arrest transition is still lacking. Particularly, the role of external boundary conditions on flow-arrest transitions, while crucial, has been not been well-characterized.

A well-known jamming phase diagram for frictionless particles predicts that a potentially flowing state can be arrested by increasing its solid volume fraction $\phi$ or decreasing its internal shear stress $\tau$ across an yield stress line.\cite{liu1998} The scenario is more complicated for frictional particles with the emergence of solid-like fragile and shear-jammed states at low volume fractions.\cite{bi2011} In practice however, the volume fraction of granular materials is rarely controlled and the material responds to external stresses by compacting under pressure, and compacting or dilating under shear.\cite{kabla2009} The dynamics of granular materials under controlled pressure $p$ are better understood through dimensionless $\mu(I)$ and $\phi(I)$ constitutive relationships, where $\mu=\tau/p$ is a dimensionless stress ratio, and $I=\dot{\gamma}d\sqrt{\rho/p}$ is a dimensionless shear rate or inertial number.\cite{forterre2008} Here $\dot{\gamma}$ is the strain rate of flow, $d$ is the mean particle size and $\rho$ is the particle density. Previous experiments\cite{boyer2011a,clavaud2017} and simulations\cite{dacruz2005,peyneau2008} have demonstrated that granular materials flow only when $\mu$ exceeds a critical value, $\mu_{c}$, along with the material attaining a critical volume fraction $\phi_{c}$. These critical conditions depend significantly on interparticle friction, as shown previously in simulations \cite{srivastava2019b} and also predicted by the critical state theory of soil mechanics.\cite{schofield1968,rothenburg2004} However, these constitutive relationships do not predict the state of the flowing granular material that eventually arrests when $\mu$ is reduced below $\mu_{c}$. Furthermore, these relationships are isotropic and do account for dominant microstructural features such as directional force chains that are observed in experiments at the flow-arrest transition.\cite{bi2011}

\begin{figure}[t]
\centering
\includegraphics{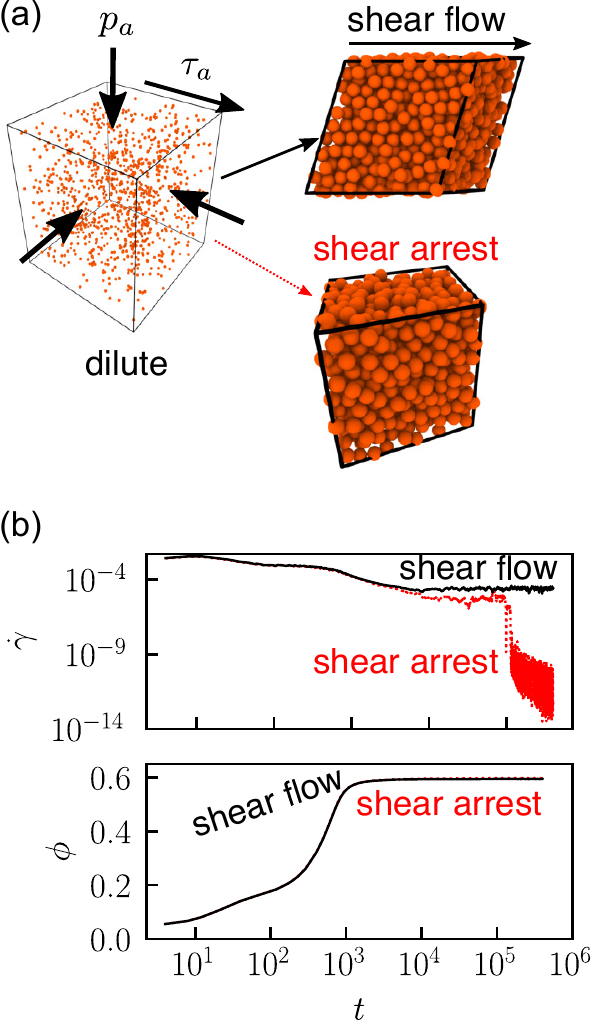}
\caption{(a) Schematic of the simulation method. The left image depicts a starting dilute state. The black lines enclosing the particles denote triclinic periodic boundaries. At $t=0$ the system is subjected to a constant external pressure $p_a$ and shear stress $\tau_a$. The images to the right depict two possible steady states at long times: shear arrest and steady shear flow for low and high values of $\tau_a/p_a$ respectively. Visually, the flowing and arrested states are indistinguishable. (b) To distinguish between the state of shear arrest (dotted red line) and steady shear flow (solid black line), the evolution of the strain rate $\dot{\gamma}$ (top) is monitored with simulation time $t$, where distinct behaviors are observed, even though the evolution of volume fraction $\phi$ with time (bottom) is similar in the two cases (Note: the two curves are almost coincident).
}
\label{fig1}
\end{figure}

Boundary conditions play a crucial role in the rheological behavior of granular materials. Under a constant external pressure and applied strain rate, $\mu$ increases monotonically with $I$ and saturates at a critical value $\mu_c$ as $I\to0$ in the quasi-static regime, whereas $\phi$ decreases monotonically with increasing $I$ reaching a critical value $\phi_{c}$ as $I\to0$.\cite{srivastava2019b} The rheological response is significantly different when the material is not allowed to dilate or compact in response to shear flow at a constant volume fraction. In this case, below $\phi_{c}$, the material can flow at all strain rates exhibiting a continuous transition from plastic to inertial flow as the strain rate is increased, whereas a solid-like yield behavior is observed above $\phi_{c}$.\cite{otsuki2011,ciamarra2011} However, the transition from fluid-like to solid-like states is discontinuous, which leads to chaotic dynamics \cite{grob2016} and a re-entrant jamming transition near $\phi_{c}$.\cite{fall2008} The situation becomes increasingly complex in some non-Brownian suspensions where external stresses can drive a frictional transition within the particle contact network \cite{wyart2014} leading to a discontinuous increase in the suspension viscosity and shear jamming.\cite{brown2012,peters2016} The nature of boundary conditions dominantly governs the mechanics of such suspensions, as demonstrated by intriguing flow phenomena such as negative dynamic compressibility\cite{dong2017} and vorticity banding.\cite{rathee2020} 

As such, a careful characterization of the flow-arrest transition as an intrinsic \emph{bulk} property of granular materials requires three key considerations of the boundary conditions: (i) a constant external pressure as the natural boundary condition, where the material can dilate or compact as it flows or arrests. In addition to being the predominant boundary condition in practical applications of granular flows, constant pressure conditions are particularly well-suited for exploring granular flow dynamics near the critical jamming volume fraction where the magnitude of stress fluctuations can be quite large;\cite{kawasaki2015,srivastava2019}~(ii) a constant external stress rather than strain rate as the imposed driving force, which allows for a seamless transition between flowing and arrested states of granular materials;\cite{wang2015,srivastava2019}~(iii) the avoidance of external walls or boundaries that can complicate the rheological response by flow localization\cite{schall2010} and non-local effects.\cite{kamrin2015}

In this work, we use stress-controlled discrete element simulations to analyze steady states of shear flow and shear arrest. Starting from a dilute state, we simulate steady shear flow and shear arrest along the paths of constant external pressure and shear stress, and identify the critical flow-arrest transition for several interparticle frictions. A state of shear arrest occurs if the applied shear stress (for a given applied pressure) is not large enough, whereas when the applied shear stress is larger than a critical value, the granular material flows steadily. We clearly identify and distinguish shear-arrested and steady flowing states along $\phi-\mu$ and $Z-\mu$ axes, where $Z$ is the coordination number. We demonstrate that the internal state of steady shear flows is uniquely represented by either $\mu$, $\phi$ or $Z$ through relationships such as $\mu(I)$ and $\phi(I)$, where we observe good agreement with existing results from rate-controlled simulations of granular flows. However, for states that eventually arrest for sub-critical applied shear stress, such uniqueness is very weak and these states are better distinguished by a higher-order structural description of the particle contact network. We identify the contact fabric tensor as an important higher-order structural descriptor that uniquely represents the steady states of both shear arrest and shear flow in granular materials. These findings have important consequences for constitutive modeling of granular materials across their fluid-like and solid-like states of existence.

\section{Model and Methods}
We simulate flow-arrest transition in granular materials by subjecting a dilute system of particles to constant external pressure $p_{a}$ and shear stress $\tau_{a}$. The particles are initially contained in a cubic simulation cell that is periodic along all directions, as shown in Fig.~\ref{fig1}(a). The stress-controlled simulation method allows for the dynamical evolution of all the shear degrees of freedom and the volume of the simulation cell in response to applied shear stress and pressure. The deformation of the simulation cell is tracked by the time evolution of its triclinic periodic boundaries denoted by a matrix $\mathbf{H}$ that is a concatenation of the three unit cell vectors that defines the system periodicity. Under the action of applied stress, the cell can dilate or compact, and shear in all possible ways, thus simulating the true bulk response of the granular material. The reader is referred to ref.\cite{srivastava2019b} for a detailed description of the simulation method. The technical details associated with the method are also provided in the Appendix.

The motion of the triclinic simulation cell $\mathbf{H}$ results in a bulk velocity gradient $\nabla \mathbf{v} = \dot{\mathbf{H}}\mathbf{H}^{-1}$ from which a symmetric strain rate tensor is computed as $\mathbf{D}=\frac{1}{2}\left(\nabla \mathbf{v} + \nabla \mathbf{v}^{T}\right)$. In steady state, $\mathbf{D}$ is trace-free, and the strain rate magnitude is computed as $\dot{\gamma}=\sqrt{\frac{1}{2}\mathbf{D}:\mathbf{D}}$.\cite{srivastava2019b} The internal Cauchy stress tensor of the system $\mathbf{\sigma}$ is computed from the dyadic product of the pairwise contact force $\mathbf{f}_{c}$ between particles with center-to-center contact vector $\mathbf{r}_{c}$ as $ \mathbf{\sigma} = (1/V)\sum_{N_c} \mathbf{f}_{c} \otimes \mathbf{r}_{c}$, where the sum is over all $N_c$ contacts, and $V$ is the volume of the parallelepiped simulation cell. The kinetic contribution to the internal stress is minimal in the dense flows considered here and is ignored. The internal pressure $p=\frac{1}{3}\sum_{i}\sigma_{ii}$, and the magnitude of internal shear stress $\tau = \sqrt{\frac{1}{2}\mathbf{\tau_D}:\mathbf{\tau_D}}$, where $\mathbf{\tau_D} = \mathbf{\sigma}-p\mathbf{I}$, and $\mathbf{I}$ is the identity matrix.

Each simulation consists of $N=10^{4}$ spherical particles whose diameters are uniformly distributed between $0.9d$ and $1.1d$. The particles interact through a linear spring-dashpot viscoelastic contact mechanical model, along with tangential Coulomb friction that is characterized by a coefficient of friction $\mu_s$.\cite{srivastava2019b} The tangential spring stiffness is set equal to the normal spring stiffness $k_n$, which is set to unity. The normal velocity damping constant is set as $\nu_n=0.5$, and the tangential velocity damping constant is set as $\nu_t=0.25$. In the present simulations, time is normalized by a characteristic timescale $t_c=\pi\left(2k_n/\rho d^{3}-\nu_n^{2}/4\right)^{-1/2}$, which is the characteristic collision time between two particles.\cite{srivastava2019b}.The simulation time step is set to $0.02t_c$, and each simulation is run for at least $10^{6}t_c$ total time. The spring constant sets the scale for stress in the system; therefore, all stresses are scaled by $k_n/d$. All the stress-controlled simulations are  performed  using  the  large-scale  molecular  dynamics  software LAMMPS.\cite{plimpton1995}

\section{Results}
The granular system initially responds to the applied $p_a$ and $\tau_a$ by rapid compaction under the action of pressure along with significant shear straining, as shown by the evolution of deviatoric strain rate $\dot{\gamma}$ and volume fraction $\phi$ in Fig.~\ref{fig1}(b). After initial transients the system enters into a quasi-steady flow that is characterized by fluctuations around a mean value of $\dot{\gamma}$ and $\phi$. At long times, the dynamical evolution of the granular system exhibits two distinct phenomena as shown in Fig.~\ref{fig1}(b): (i) for low values of $\tau_{a}/p_{a}$, the system enters into a dynamically arrested solid-like state, as evidenced by a drop of $\dot{\gamma}$ by several orders of magnitude followed by a slow creeping deformation;\cite{srivastava2017} (ii) for large enough values of $\tau_{a}/p_{a}$ the system continues to flow steadily around mean values of $\dot{\gamma}$ and $\phi$.

\begin{figure}[t!]
\centering
\includegraphics{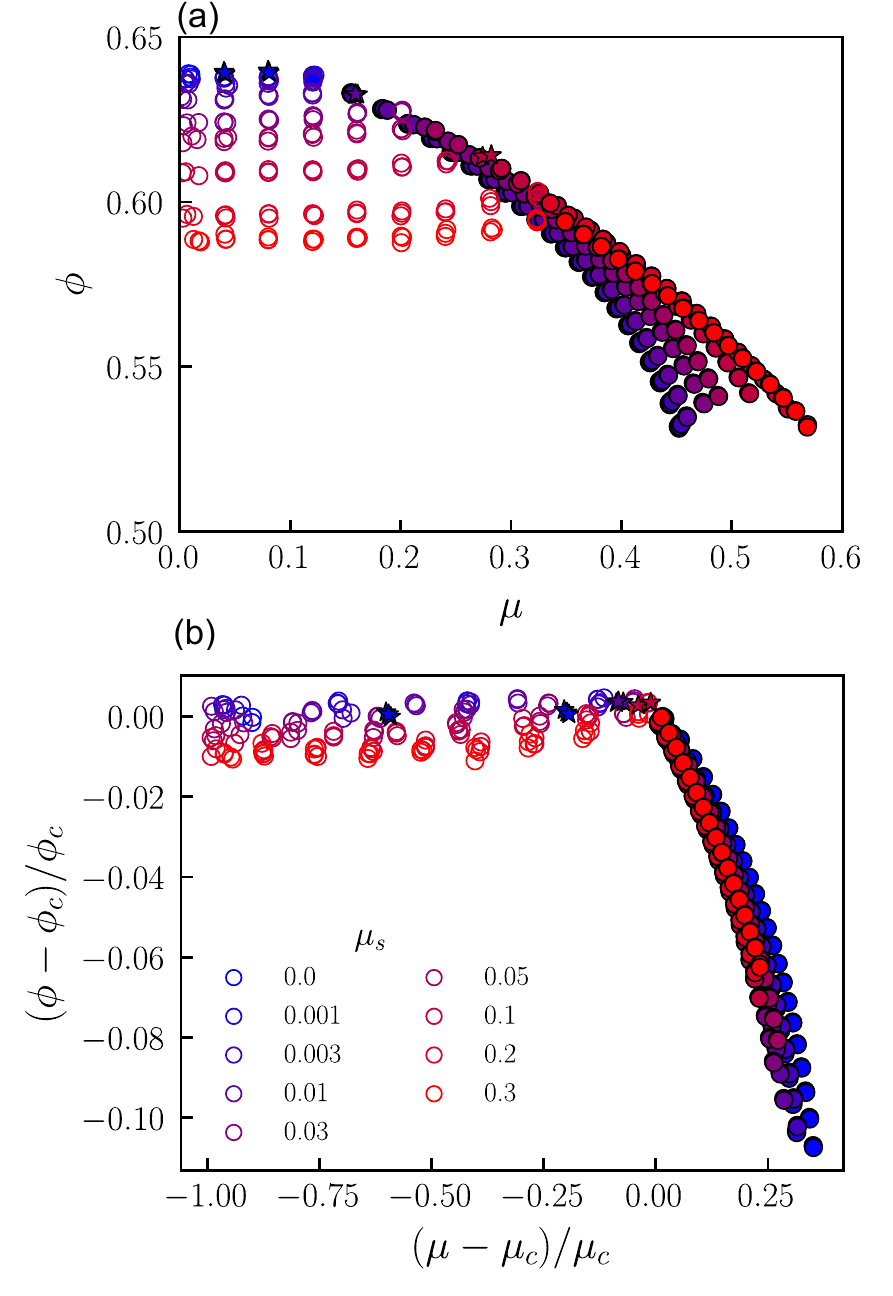}
\caption{(a) All simulated steady states of shear arrest (open symbols) and shear flow (closed symbols) on $\phi-\mu$ axes for all simulated interparticle frictions $\mu_{s}$ (see the color legend in (b)). The asterisk denote steady states in the vicinity of the flow-arrest transition for which some simulations arrested and some steadily flowed at long times. (b) All the simulated steady states of shear arrest and shear flow in (a), with each shifted and normalized by their $\mu_s$-dependent critical values $\mu_{c}$ and $\phi_{c}$.}
\label{fig2}
\end{figure}

Previously it was demonstrated that the dynamical arrest of dense granular flows is highly stochastic, and the time for the flow to arrest exhibits a heavy-tailed distribution whose statistics diverge at a $\mu_s$-dependent critical value of the stress ratio $\mu_c$.\cite{srivastava2019} However, the internal microstructural state of the granular material upon arrest is deterministic.\cite{srivastava2020} Based on these observations, we postulate that steady states of arrest exist at low $\mu$ and high $\phi$, and states of steady flow exist at high $\mu$ and low $\phi$, with $\mu_s$-dependent critical values of $\mu_c$ and $\phi_c$ bifurcating the states of flow and arrest. Furthermore, it is expected that no steady state can exist at high $\mu$ and high $\phi$, or low $\mu$ and low $\phi$. For $\mu$ greater than $\mu_c$, the system cannot exist at arbitrarily high $\phi$ and will necessarily dilate to a volume fraction lower than $\phi_c$ to achieve steady flow. Similarly, for $\mu<\mu_c$, the shear stress is not large enough to drive steady granular flow and the material will compact into a shear-arrested solid under the action of external pressure. For the particular case of $\mu=0$, i.e., at a finite external pressure and zero shear stress, the system will evolve towards a $\mu_s$-dependent isotropic-jammed state at volume fraction $\phi_{J}$.\cite{silbert2010,santos2020} Although the steady states of shear arrest and shear flow calculated in this paper are extracted from simulations starting from states with very low volume fractions, we have verified that they are robust to initial conditions by performing simulations with initial states at higher volume fractions, but still lower than $\phi_c$.

\subsection{States of Flow and Arrest}
Simulations were performed for interparticle frictions ranging from $\mu_s=0.0$ for frictionless particles to $\mu_s=0.3$ that characterizes the high friction limit. Several applied stresses ranging from $\tau_a/p_a=0.0$ for isotropic jamming to $\tau_a/p_a=1.0$ were analyzed, and three simulations were run for each case of $\mu_s$ and $\tau_a/p_a$. Although the results presented in the main text are for $p_a=10^{-5}$, which corresponds to the hard-particle limit, no significant pressure dependence was observed for pressures $p_a=10^{-4}$ and $p_a=10^{-6}$ in accordance with similar previous observations\cite{favierdecoulomb2017,srivastava2019b} (steady states of shear arrest and flow corresponding to these other two pressures are included in the ESI$^\dag$).
Figure~\ref{fig2}(a) shows all the simulated steady states of shear arrest (open symbols) and shear flow (closed symbols) for all the simulated $\mu_s$ on $\phi-\mu$ axes. The steady state behavior for a given case of $\mu_s$ and $\tau_a/p_a$ is treated as shear arrest (flow) if all three simulations for that case resulted in shear arrest (flow). In the vicinity of critical transition between arrest and flow, often not all three simulations resulted in shear arrest or flow within the simulation run time; these cases have been marked with an asterisk. This results from the stochastic nature of shear arrest and depends significantly on simulation run time and system size.\cite{srivastava2019} The empty region at low $\phi$ and low $\mu$, and high $\phi$ and high $\mu$ in Fig.~\ref{fig2}(a) is inaccessible at steady state.

For steady flowing states, $\phi$ decreases rapidly and monotonically with increasing $\mu$ for all $\mu_s$, thus indicating the dilating nature of granular flows. Furthermore, the one-to-one relationship between $\phi$ and $\mu$ implies that $\mu$ uniquely sets $\phi$ in steady granular flows for all $\mu_s$, which has important consequences in constitutive modeling of granular materials.\cite{kim2020} This is similar to the predictions of the kinetic theory\cite{lun1984a} and hydrodynamic models\cite{bocquet2001} where such a relationship was determined between internal granular temperature and $\phi$ for granular flows. Unlike states of steady flow, a strong one-to-one relationship between $\mu$ and $\phi$ does not exist for the states of shear arrest, which all seem to possess nearly the same $\mu_s$-dependent $\phi$ irrespective of $\mu$. A slight increase in $\phi$ with $\mu$ is observed near the flow-arrest transition for particles with high friction, and this can possibly be attributed to rheological hysteresis in frictional particles.\cite{degiuli2017} Therefore, another internal variable beyond $\phi$ is required for a unique characterization of the internal state of shear arrest. 

\begin{figure}[t!]
\centering
\includegraphics{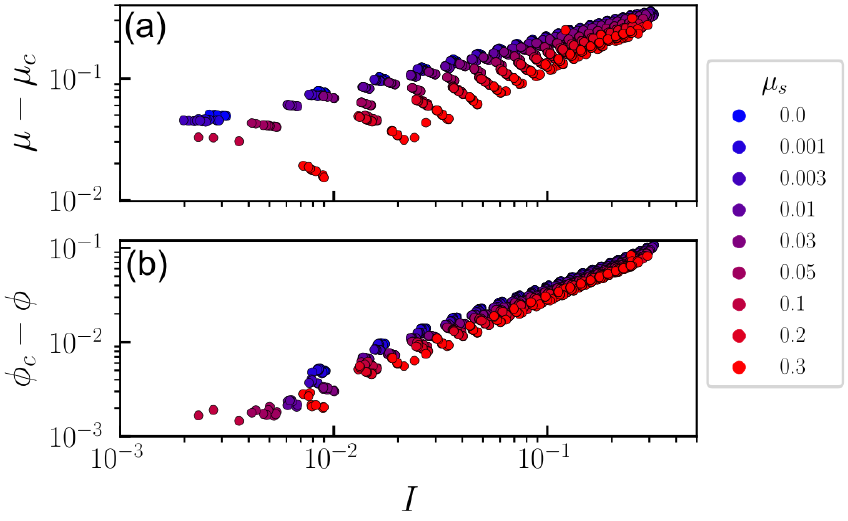}
\caption{Variation of (a) $\mu-\mu_c$ and (b) $\phi_c -\phi$ with the inertial number $I$ for simulated states of steady shear flow. The different colors represent various interparticle frictions $\mu_s$ (see legend).}
\label{fig3}
\end{figure}

Considering closely the region of arrested states from our simulations, the isotropic jamming volume fraction (corresponding to $\mu=0$) for frictionless particles $\phi_J=0.64$ is equivalent to the random closed packing fraction, whereas $\phi_J=0.59$ is observed for high friction, similar to previous simulations\cite{silbert2010} and experiments\cite{jerkins2008,farrell2010,boyer2011a,clavaud2017} on frictional particles. The critical flow-arrest transition for frictionless particles occurs at $\phi_c=0.64$ and $\mu_c=0.1\pm0.02$ (denoted by the green dot in the schematic in Fig.~\ref{fig1}(c)), which is consistent with previous simulations\cite{peyneau2008} and experiments\cite{clavaud2017} on the rheology of frictionless particles. The equality $\phi_J = \phi_c = 0.64$ with the random closed packing fraction for frictionless particles confirms previous observations that frictionless particles do not need to dilate in order to begin flowing \cite{peyneau2008}. For particles with high friction, the critical flow-arrest transition occurs at $\phi_c=0.59$ and $\mu_c=0.34\pm0.01$  (denoted by the blue dot in the schematic in Fig.~\ref{fig1}(c)), which is consistent with previous experiments,\cite{boyer2011a,clavaud2017} simulations,\cite{salerno2018,srivastava2019b} and predictions from the theory of critical state soil mechanics\cite{schofield1968,rothenburg2004} for the onset of granular flow. Although $\mu_c$ and $\phi_c$ were estimated from the discrete data by demarcating the steady shear flow and shear arrested states, these estimates match well with their more precisely calculated values from the power-law divergence of the time for a flowing granular material before arrest.\cite{srivastava2019} Using the estimated values of $\mu_c$ and $\phi_c$, all the states of shear arrest for all $\mu_s$ collapsed on to a nearly horizontal master curve along $(\phi-\phi_c)/\phi_c$ vs. $(\mu-\mu_c)/\mu_c$ axes, as shown in Fig.~\ref{fig2}(b). The very weak dependence of $\phi-\phi_c$ on $\mu-\mu_c$ for shear-arrested states also demonstrates that $\phi_J\approx\phi_c$, i.e., granular materials begin to flow at volume fractions that are nearly equal to their friction-dependent isotropic jamming volume fractions.

Unlike shear arrest, the states of steady flow do not collapse onto a similar master curve because $\mu-\mu_c$ and $\phi-\phi_c$ scale differently with the inertial number $I$ of the flow for different interparticle frictions, as shown in Fig.~\ref{fig3}. For a shear flow in steady state, the $\mu(I)$ and $\phi(I)$ rheology obtained from the stress-controlled simulations should be consistent with similar scalings obtained from rate-controlled simulations. Here we confirm that the $\mu_s$-dependent power-law scaling of $\mu-\mu_c \sim I^{\alpha}$ and $\phi_c-\phi \sim I^{\beta}$ obtained from the present stress-controlled simulations (also described in detail in ref.\cite{srivastava2019b}) correspond well with rate-controlled simulations reported in ref.\cite{favierdecoulomb2017} and theoretical predictions.\cite{degiuli2016}

\subsection{The Role of Coordination and Contact Fabric}

\begin{figure}[t!]
\centering
\includegraphics{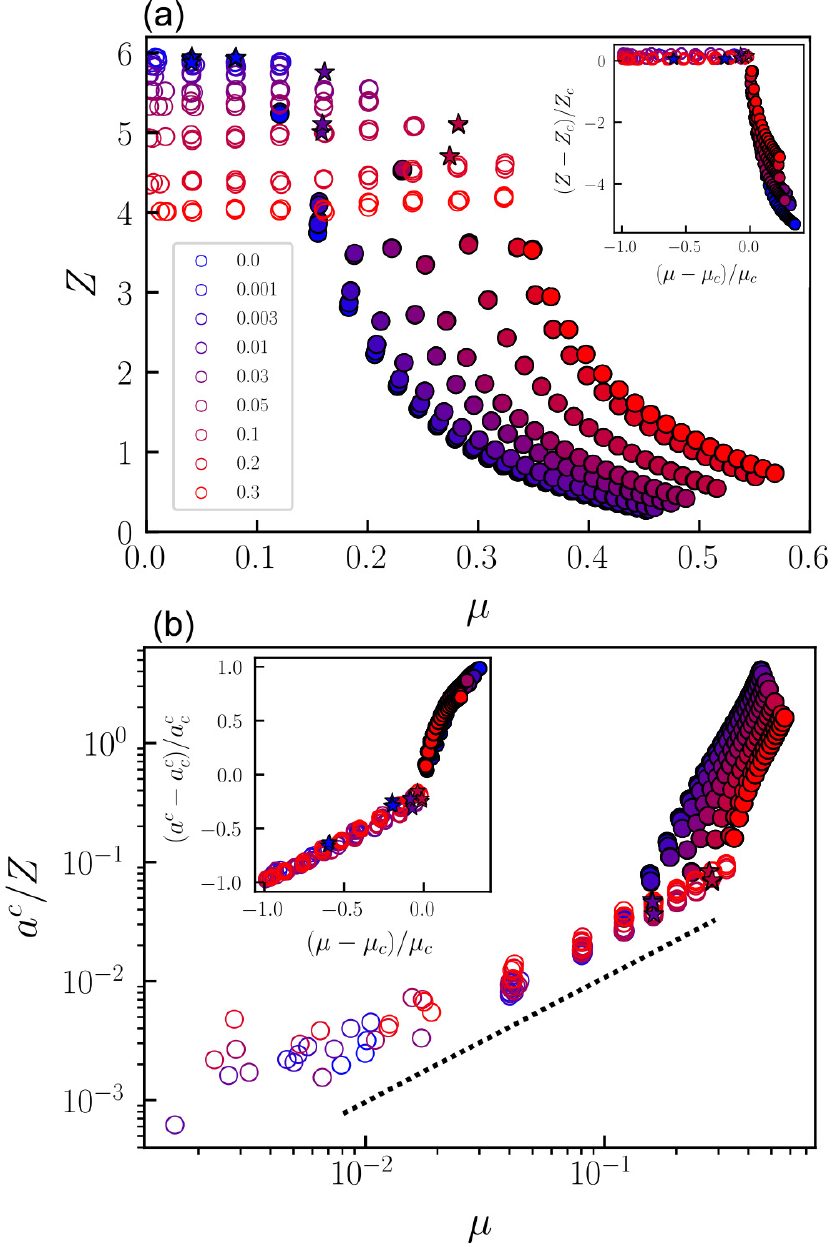}
\caption{(a) All simulated steady states of shear arrest (open symbols) and shear flow (closed symbols) on $Z-\mu$ axes for all simulated interparticle frictions $\mu_{s}$ (see legend in (a)). The asterisk denote states in the vicinity of the flow-arrest transition for which some simulations arrested and some flowed steadily at long times. The inset shows all these states shifted and normalized by their $\mu_s$-dependent critical values $Z_c$ and $\mu_c$. (b) Contact fabric anisotropy $a^{c}$ normalized by $Z$ for the states of steady shear flow and shear arrest as a function of $\mu$ for different $\mu_s$. The dotted line denotes a slope of unity. The inset shows all these states shifted and normalized by the $\mu_s$-dependent critical values $a^{c}_c$ and $\mu_c$.}
\label{fig4}
\end{figure}
The coordination of the grains in a granular system plays a key role in their properties both during jamming\cite{goodrich2016} and flow.\cite{sun2011,azema2014} Previously, a constitutive model was proposed that incorporated the coordination number as an internal state variable in steady and unsteady quasi-static granular flows.\cite{sun2011} In this work, we define a coordination number $Z=2N_c/N$, where $N_c$ is the total number of contacts with non-zero normal forces. In Fig.~\ref{fig4}(a), $Z$ for flow and arrested states is plotted as a function of $\mu$ for various $\mu_s$. For frictionless particles, the coordination number at isotropic jamming $Z_J\approx6$, whereas $Z_J\approx4$ for particles with high friction, which is consistent with previous simulations.\cite{silbert2010,santos2020} In a manner similar to $\phi$, all the arrested states possess the same $\mu_s$-dependent $Z$ irrespective of $\mu$. Upon demarcating the steady states of shear arrest and shear flow, critical values of $\mu_s$-dependent $Z_c$ were extracted, and all the shear-arrested states are collapsed on to a nearly horizontal master curve for all $\mu_s$ by plotting $(Z-Z_c)/Z_c$ vs. $(\mu-\mu_c)/\mu_c$, as shown in the inset of Fig.~\ref{fig4}(a). For flowing states, $Z$ decreases rapidly and monotonically with $\mu$ for all $\mu_s$, predominantly from the loss of contacts in the extension direction of simple shear flow.\cite{radjai2012} For all steady flowing states, a unique one-to-one relationship exists between $Z$ and $\mu$ as shown in Fig.~\ref{fig4}(a), in a manner similar to the relationship between $\phi$ and $\mu$ in Fig.~\ref{fig2}(a). Therefore for a given $\mu_s$, the steady shear flowing state of a granular material can be uniquely identified by either $\mu$, $\phi$ or $Z$, as there exists a one-to-one relationship between these quantities.

Such a one-to-one relationship is very weak for steady states that eventually arrested because of insufficient applied shear stress. These states possess nearly the same $\mu_s$-dependent $\phi$ and $Z$ regardless of $\mu$, thus necessitating a more microstructure-sensitive metric for their characterization. It is expected that the topological structure of the particle contact network in an arrested granular system would be increasingly anisotropic as $\mu$ increases, resulting from directionally-dominant contact networks that are required to support the external shear stress.\cite{bi2011,srivastava2020} Therefore, a higher-order structural descriptor beyond isotropic measures such as $\phi$ and $Z$ is required to distinguish structural anisotropy in shear-arrested states. We quantify the structural anisotropy using a contact fabric tensor $\mathbf{A}^c$, which provides a convenient description of the directional distribution of the particle contact network.\cite{radjai1998,srivastava2020} It can be defined as the coefficient of the second-order Fourier expansion of the orientational distribution function $P(\mathbf{n})$ of unit vectors $\mathbf{n}$ connecting two contacting particles, such that $P(\mathbf{n})=\frac{1}{4\pi}\left[1+\mathbf{A}^c:\left(\mathbf{n}\otimes\mathbf{n}\right)\right]$. The tensor $\mathbf{A}^c$ is trace-free and symmetric, and its magnitude $a^c = \sqrt{\frac{1}{2}\mathbf{A}^c:\mathbf{A}^c}$ provides a measure of the anisotropy in the contact network. Figure~\ref{fig4}(b) shows the dimensionless ratio $a^c/Z$ for flowing and arrested states as a function of $\mu$ for various $\mu_s$. Unlike $\phi$ and $Z$, there exists a one-to-one relationship between $a^c$ and $\mu$ for the states of shear arrest. Furthermore, a linear relationship $a^c\sim\mu$ emerges as $Z$ is nearly constant for all shear arrested states regardless of $\mu$ from Fig.~\ref{fig4}(a). Interestingly, the linear relationship between $a^c/Z$ and $\mu$ is independent of $\mu_s$ (see the collapse of $a^c/Z$ for all $\mu_s$ in Fig.~\ref{fig4}(b)), thus indicating that the internal state of a shear-arrested granular system containing weakly polydisperse spheres is uniquely characterized by its contact anisotropy irrespective of the friction. A critical $\mu_s$-dependent fabric anisotropy $a_{c}^{c}$ at the flow-arrest transition is extracted by demarcating the steady states of shear arrest and shear flow. The unique linear relationship between fabric anisotropy and stress ratio for all shear-arrested states regardless of $\mu_s$ is also seen by the collapse of $(a^c-a_{c}^{c})/a_{c}^{c}$ vs. $(\mu-\mu_c)/\mu_c$ on to a linear master curve in the inset of Fig.~\ref{fig4}(b).

For steady flowing states above the critical flow-arrest transition, the ratio $a^{c}/Z$ increases rapidly with $\mu$ as shown in Fig. \ref{fig4}(b), with an observed scaling of $a^c/Z \sim \mu^{\xi}$, where $\xi$ varies slightly with $\mu_s$ ranging from $3.8$ for frictionless particles to $4.0$ for particles with high friction. Unlike shear arrested states where there is no loss of coordination upon increasing $\mu$, the super-linear increase in $a^c$ with $\mu$ in flowing states is a consequence of directional alignment of the contacts along the compression direction in addition to a loss of coordination in the extension direction of shear flow.\cite{radjai2012} As a result, there is a discontinuity in the variation of $a^c/Z$ with $\mu$ at the flow-arrest transition, as also seen in the inset of Fig. \ref{fig4}(b). Such a rapid increase of contact anisotropy in high shear rate granular flows was also demonstrated previously.\cite{azema2014} Therefore, while $\phi$ or $Z$ can completely characterize the steady state of shear flow in granular materials, $a^c/Z$ is a key constitutive variable that uniquely characterizes both steady states of shear arrest and shear flow in granular materials that are driven by a constant applied shear stress and pressure from an initially dilute state, and is remarkably insensitive to the applied pressure (see ESI$^\dag$ for critical values of  at other pressures). However, it is expected that the ability of $a^c/Z$ to completely determine a shear-arrested state will breakdown if the arrest is approached along a different path, such as shearing from an initially solid state, which is a consequence of path-dependent mechanics of granular materials in their arrested (or jammed) states.\cite{kumar2016} Furthermore, the uniqueness of $a^c/Z$ as a descriptor of steady states of both arrest and flow is also expected to breakdown for unsteady flows such as shear reversal where the fabric also evolves in time along with the mechanics of the material.\cite{sun2011,parra2019} A detailed investigation of these issues constitute an important part of the future work.

\subsection{Critical States at the Flow-Arrest Transition}
\begin{figure}[t!]
\centering
\includegraphics[width=\linewidth]{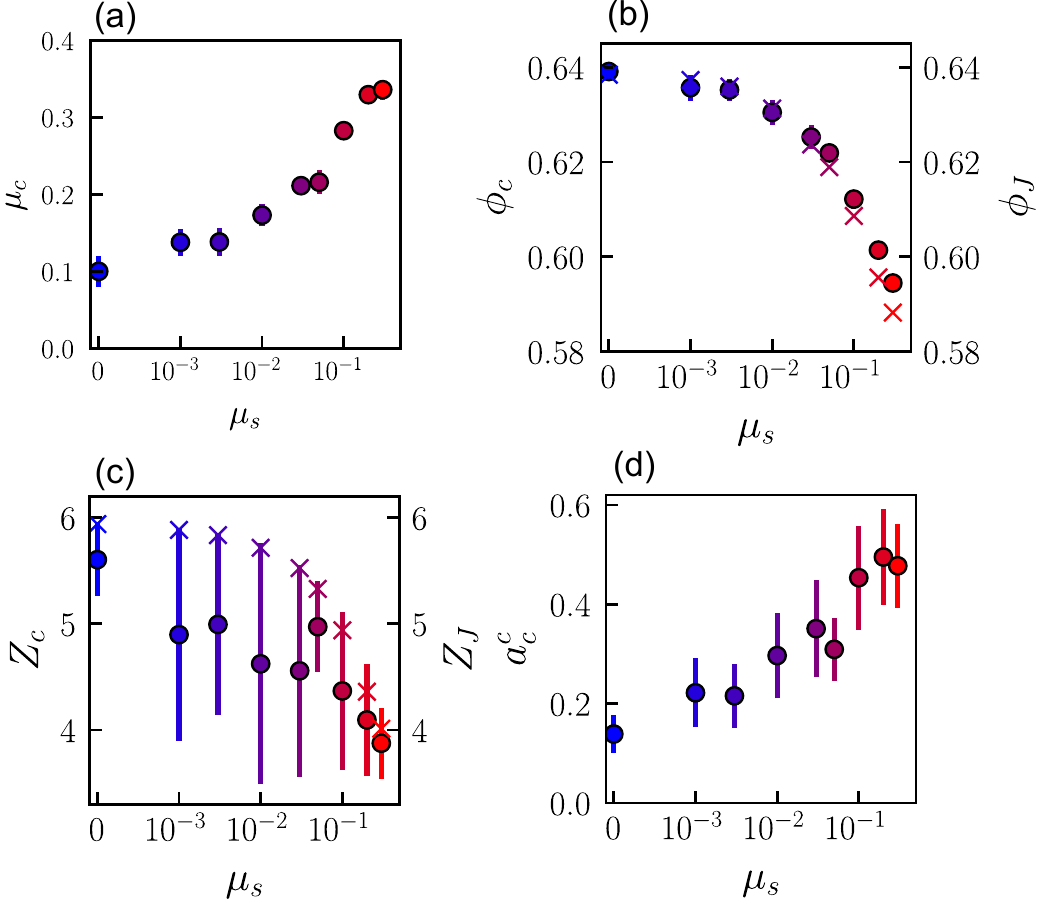}
\caption{Critical values (a) $\mu_{c}$, (b) $\phi_{c}$, (c) $Z_c$ and (d) $a_{c}^{c}$ as a function of friction $\mu_{s}$. The jamming volume fraction $\phi_J$ and coordination number $Z_J$ are also marked with crosses in (b) and (c) respectively. The leftmost data points in (a) - (d) correspond to the frictionless case. The vertical bars around data points represent the error in estimating the critical boundary between the steady shear flowing and shear arrested granular states from the discrete simulation data in Figs.~\ref{fig2}(a) and \ref{fig4}.}
\label{fig5}
\end{figure}

Simulations under controlled pressure and shear stress facilitate a seamless transition between shear-arrested and shear-flowing states of granular materials. Such simulations are well-suited for identifying the critical boundary that separates these two steady states. We extract $\mu_s$-dependent critical values of volume fraction $\phi_{c}$, stress ratio $\mu_{c}$, coordination $Z_{c}$ and contact anisotropy $a_{c}^{c}$ at the flow-arrest transition. The critical stress ratio $\mu_{c}$ varies monotonically with $\mu_{s}$, and ranges from $\mu_c=0.1\pm0.02$ for frictionless particles to $\mu_c=0.34\pm0.01$ for particles with high friction, as shown in Fig.~\ref{fig5}(a). The variation of $\mu_c$ with $\mu_s$ also corresponds well with previous simulations that more precisely characterized the critical transition from the power-law divergence of the time to arrest.\cite{srivastava2019} The critical $\mu_c$ corresponds to the dynamical arrest of granular flows, i.e., the minimum shear stress required to continue flowing a granular material indefinitely. Although this critical shear stress ratio is considered equal to the value required to start a granular flow in standard granular rheological models,\cite{forterre2008} recent experiments\cite{perrin2019} and simulations\cite{degiuli2017} have identified a mild hysteresis in granular rheology that can result in different $\mu_c$, or equivalently the distinction between the flow angle and the angle of repose in inclined plane flows,\cite{silbert2005,mowlavi2021a} depending on whether the flow-arrest boundary is reached from a flowing state or an arrested state (such as in start-up shear tests).

The critical volume fraction $\phi_c$ at the flow-arrest transition decreases monotonically with $\mu_s$, and ranges from $\phi_c=0.64$ for frictionless particles to $\phi_c=0.59$ for particles with high friction, as shown in Fig.~\ref{fig5}(b). Frictionless particles can flow at their densest random closed pack volume fraction, whereas the presence of friction necessarily requires dilation for steady flow. If the dilation of the granular material containing frictional particles is restricted and the material is forced to flow in a volume- and strain-controlled setup, as described in previous simulations\cite{heussinger2013,otsuki2011} and experiments,\cite{brown2012} the flow will be chaotic and prone to instabilities. Such frustrated dilatancy effects have also been proposed to cause discontinuous shear thickening in dense suspensions.\cite{brown2012} As shown in Fig.~\ref{fig5}(b), $\phi_c$ is nearly equal to the isotropic jamming volume fraction $\phi_J$, thus demonstrating that shear arrest of granular materials occurs at a well-defined volume fraction.

Unlike the volume fraction, a precise determination of the critical $Z_c$ at the flow-arrest boundary is challenging because the granular material loses a significant number of contacts upon flow initiation. This is demonstrated in Fig.~\ref{fig5}(c), where the critical coordination $Z_c$ is consistently lower than the isotropic jamming coordination $Z_J$, and they both decrease with $\mu_s$. For frictionless particles $Z_J=5.94$ and $Z_c=5.6\pm0.34$ whereas $Z_J=4.01$ and $Z_c=3.88\pm0.33$ for particles with high friction. The large error bars associated with $Z_c$ indicate that the loss of coordination during transition from arrested states to flowing states has significant variability across simulations, resulting from random breaking and forming of contacts during steady shear flow. 

The critical contact fabric anisotropy $a_c^{c}$ denotes the maximum anisotropy---i.e., maximum structural alignment of the contact network in the shear direction---that a granular material can sustain above which it will necessarily flow. The critical anisotropy increases with $\mu_s$ and ranges from $a_c^{c}=0.14\pm0.03$ for frictionless particles to $a_c^{c}=0.48\pm0.08$ for particles with high friction, as shown in Fig.~\ref{fig5}(d). The increase of $a_c^{c}$ with $\mu_s$ is expected as the tangential force of friction between particles provides increased stability to a highly sheared but static granular network. Furthermore, the mechanical stability of such highly sheared systems has also been recently demonstrated in experiments on shear jamming where the application of shear strain introduces directional rigidity in frictional granular systems at volume fractions below the isotropic $\phi_J$.\cite{bi2011} Shear jamming is not observed in the present simulations because the material is allowed to dilate or contract under constant pressure conditions, and a sheared granular material at low volume fractions will necessarily compact towards a shear arrested state or dilate towards steady shear flow depending on the applied stress and pressure. However, it is expected that the phenomenon of shear arrest at constant pressure simulated in this study is intimately related to the phenomenon of shear jamming at constant volume, as also hypothesized in shear jamming experiments.\cite{bi2011}

The non-zero value of $a_{c}^{c}$ for frictionless particles, as shown in Fig.~\ref{fig5}(d), highlights that even dense frictionless granular systems can sustain a limited amount of structural anisotropy without yielding to flow. In the absence of friction and in the limit of hard particle stiffness, the origins of such non-zero yield stress are purely geometrical in nature.\cite{roux2000} Furthermore, these results indicate that the phenomenon of shear jamming can be observed for frictionless particles as well. Recent computational investigations have confirmed shear jamming-like behavior in frictionless systems\cite{vinutha2016} along with connections to shear dilatancy\cite{das2020} that is ubiquitous in frictional systems but can also occur in frictionless systems.\cite{babu2021}

\section{Conclusions}
We have utilized a recently-developed, novel, stress-controlled methodology to simulate the flow-transition in granular materials for varying interparticle frictions. We have characterized steady states of shear arrest and shear flow that occur in these materials when an initially dilute configuration of the material is subjected to a constant external shear stress and pressure.  For steady flowing states, we have demonstrated that there exists a unique one-to-one relationship between internal stress ratio $\mu$, volume fraction $\phi$ and coordination number $Z$, which are characterized by the well-known $\mu(I)$ and $\phi(I)$ dimensionless relationships. In contrast, for shear arrest such one-to-one relationships break down, and an additional higher-order structural descriptor in the form of a contact fabric tensor $\mathbf{A}^c$ is required to fully characterize the steady state. Furthermore, the critical steady states that demarcate the flow-arrest transition are found to be highly sensitive to interparticle friction.

These findings have important implications towards the development of microstructure-aware constitutive models that can describe the phenomenon of flow-arrest transition in granular materials. We propose that the contact fabric anisotropy is an important state variable that connects the steady state microstructure and mechanics of granular materials in their solid-like and fluid-like state of existence across the transition, particularly when a flowing granular material abruptly arrests as a result of insufficient applied shear stress.\cite{srivastava2019} Future work will involve exploring the transient response of granular materials at the flow-arrest transition. Recent developments in constitutive modeling of granular materials have also demonstrated the viability of fabric tensor as an evolving constitutive variable\cite{goddard2014} in describing transient rheological phenomena such as shear reversal.\cite{sun2011,parra2019}

Another important implication of the flow-arrest transition is observed in discontinuous shear thickening of dense suspensions and shear jamming of granular materials, where the dynamical arrest of a flowing material is caused by a frictional transition within the contact network induced by an external applied stress.\cite{wyart2014} Our stress-controlled simulation method is particularly well-suited to analyze such phenomena by enabling a precise control of the applied stresses within the simulation. Furthermore, the ability to impose an external pressure in a fully periodic setup without boundaries allows the system to dilate or contract in response to the applied stresses, thus providing avenues to disentangle the role of external boundaries\cite{brown2012} from bulk instabilities\cite{wyart2014} in these rheological transitions.

We have focused exclusively on monodisperse granular materials in simple shear, and in which sliding friction is the sole tribological phenomena. However, recent simulations have indicated that additional frictional modes such as rolling friction,\cite{santos2020} particle shape,\cite{salerno2018}, size dispersity,\cite{srivastava2021b} and the loading geometry\cite{clemmer2021} can have a dominant impact on granular jamming and rheology. Investigating flow-arrest transition for more complex particle characteristics, contact mechanics, and flow geometries beyond simple shear is an important direction for research that will provide better predictive capabilities in practical applications of granular materials.






\section*{Appendix: Stress-Controlled Simulation Methodology}
Here we provide technical details about the stress-controlled methodology that was used in this work to simulate the flow-arrest transition. The numerical method employed here has been described in detail in a previous publication.\cite{srivastava2019b}

A collection of $N$ particles is contained within a 3D triclinic simulation cell that dynamically evolves under the application of the total external stress tensor $\mathbf{\sigma}_{\mathrm{a}}$, which includes the hydrostatic contribution $p_{\mathrm{a}}$. Under the application of external stress, the simulation cell is allowed to change its volume and shear in all directions, thus simulating the bulk response of the granular material. The simulation cell remains a parallelepiped throughout the simulation, and is represented by a matrix $\mathbf{H}$, which is a concatenation of the three lattice vectors that define the periodicity of the system. The cell matrix is constrained to be upper-triangular and the internal stress tensor is symmetrized to prevent any spurious cell rotations. The equations of motion for $N$ particle positions and momenta $\{\mathbf{r}_{i},\mathbf{p}_{i}\}$, and the simulation cell matrix and its associated momentum tensor $\{\mathbf{H},\mathbf{P}_{g}\}$ are given by:
\begin{eqnarray}
\dot{\mathbf{r}}_{i}&=&\frac{\mathbf{p}_{i}}{m_{i}}+\frac{\mathbf{P}_{g}}{W_{g}}\mathbf{r}_{i},\\
\dot{\mathbf{p}}_{i}&=&\mathbf{f}_{i}-\frac{\mathbf{P}_{g}}{W_{g}}\mathbf{p}_{i}-\frac{1}{3N}\frac{\mathrm{Tr\left[\mathbf{P}_{g}\right]}}{W_{g}}\mathbf{p}_{i},\\
\dot{\mathbf{H}}&=&\frac{\mathbf{P}_{g}}{W_{g}}\mathbf{H},\\
\dot{\mathbf{P}_{g}}&=&V\left(\mathbf{\sigma}-\mathbf{I}p_{\mathrm{a}}\right)-\mathbf{H}\boldsymbol{\Sigma}\mathbf{H}^{T},
\end{eqnarray}
where $\mathbf{f}_{i}$ is the net force on a particle $i$ of mass $m_i$, $V$ is the variable volume of the simulation cell, $\mathbf{I}$ is the identity tensor, and $W_{g}$ is a `fictitious' mass associated with the inertia of the simulation cell. The tensor $\mathbf{\sigma}$ represents the internal Cauchy stress, and the tensor $\boldsymbol{\Sigma}$ is defined as:~\citep{shinoda2004}
\begin{equation}
\boldsymbol{\Sigma}=\mathbf{H}_{0}^{-1}\left(\mathbf{\sigma}_{\mathrm{a}}-\mathbf{I}p_{\mathrm{a}}\right)\mathbf{H}_{0}^{T^{-1}},
\end{equation}
where $\mathbf{H}_{0}$ is some reference state of the simulation cell, and $J^{-1}\mathbf{H}\boldsymbol{\Sigma}\mathbf{H}^{T}$ represents the `true' measure of the external deviatoric stress, which is defined with respect to the reference state.~\citep{souza1997} Here $J=\mathrm{det}\left[\mathbf{F}\right]$ is the Jacobian of the deformation gradient $\mathbf{F}$, which is defined as:
\begin{equation}
    \mathbf{F}=\mathbf{H}\mathbf{H}_{0}^{-1}.
\end{equation}
In the present stress-controlled simulation methodology, the reference configuration is updated every time step so that the $\mathbf{\sigma}$ does not deviate from $\mathbf{\sigma}_{\mathrm{a}}$ significantly. This is equivalent to resetting $\mathbf{F}$ to an identity matrix every time step. Under these conditions, after a sufficiently long simulation of the arrested state, $\mathbf{\sigma}$ is balanced by $\mathbf{\sigma}_{\mathrm{a}}$ within some numerical precision.\cite{smith2014} For steady flowing states, $\mathbf{\sigma}$ will be different from $\mathbf{\sigma}_{\mathrm{a}}$, and those differences become larger at higher strain rates.

We set the fictitious cell mass as $W_{g}=Nk_{n}d^{2}/\omega_{g}^{2}$,\cite{srivastava2019b} in analogy with similar recommendations for molecular systems.\cite{martyna1996} The choice of the damping parameter $\omega_{g}$ controls the magnitude of stress fluctuations during the simulation. A small value of $\omega_{g}$ results in larger stress fluctuations, whereas a large value of $\omega_{g}$ results in longer simulation times to achieve steady state. We choose $\omega_{g}=2.2\sqrt{m/k_{n}}$ as a reasonable value for damping, where $m$ is the mean particle mass. We have verified that $\omega_{g}$ modulates the range of inertial numbers $I$ accessible in the stress-controlled simulations, but does not change the steady flow or arrest response of the system. A comprehensive numerical analysis of the effect of $\omega_{g}$ on stress-controlled simulations of granular flows is a part of our ongoing investigation.

\section*{Conflicts of interest}
There are no conflicts to declare.

\section*{Acknowledgements}
I. Srivastava acknowledges support from the U.S. Department of Energy (DOE), Office of Science, Office of Advanced Scientific Computing Research, Applied Mathematics Program under contract No. DE-AC02-05CH11231. This work was performed at the Center for Integrated Nanotechnologies, a U.S. DOE and Office of Basic Energy Sciences user facility. Sandia National Laboratories is a multimission laboratory managed and operated by National Technology and Engineering Solutions of Sandia, LLC, a wholly owned subsidiary of Honeywell International, Inc., for the U.S. DOE’s National Nuclear Security Administration under Contract No. DE-NA-0003525. This paper describes objective technical results and analysis. Any subjective views or opinions that might be expressed in the paper do not necessarily represent the views of the U.S. DOE or the United States Government.



\balance


\bibliography{main} 
\bibliographystyle{rsc} 

\end{document}